\documentclass[aps,prd,10pt,notitlepage,nofootinbib]{revtex4-1}

\usepackage[utf8]{inputenc}
\usepackage{amsmath,amssymb,amsfonts}

\usepackage{newpxtext,newpxmath}

\usepackage{bbold}
\usepackage{bm}
\usepackage{graphicx}
\usepackage[usenames,dvipsnames]{xcolor}
\usepackage{color}
\usepackage[colorlinks=true,linkcolor=blue,urlcolor=magenta,citecolor=magenta]{hyperref}

\usepackage{slashed}
\usepackage[english]{babel}
\usepackage{dcolumn}
\usepackage{pifont}
\usepackage{dsfont,mathrsfs}
\usepackage{cancel}
\usepackage{bigints}
\usepackage{accents}
\usepackage{soul}
\usepackage{multirow}
\usepackage{natbib}
\usepackage{tikz-feynman}

\usepackage{amsmath, amssymb}
\usepackage{bbold}
\usepackage{makecell}
\usepackage{simpler-wick}
\usepackage[export]{adjustbox}

\usepackage[bb=boondox]{mathalpha}

\usepackage{orcidlink} 

\DeclareMathOperator*{\SumInt}{%
	\mathchoice%
	{\ooalign{$\displaystyle\sum$\cr\hidewidth$\displaystyle\int$\hidewidth\cr}}
	{\ooalign{\raisebox{.14\height}{\scalebox{.7}{$\textstyle\sum$}}\cr\hidewidth$\textstyle\int$\hidewidth\cr}}
	{\ooalign{\raisebox{.2\height}{\scalebox{.6}{$\scriptstyle\sum$}}\cr$\scriptstyle\int$\cr}}
	{\ooalign{\raisebox{.2\height}{\scalebox{.6}{$\scriptstyle\sum$}}\cr$\scriptstyle\int$\cr}}
}

\newcommand{\nn}{\nonumber}

\newcommand{\MB}[1]{\left|#1\right|}
\newcommand{\FB}[1]{\left(#1\right)}
\newcommand{\SB}[1]{\left\{#1\right\}}
\newcommand{\TB}[1]{\left[#1\right]}

\newcommand{\munu}{{\mu\nu}}

\newcommand{\fsl}{\slashed}

\newcommand{\unit}{\mathds{1}}  

\newcommand{\half}{\dfrac{1}{2}}
\newcommand{\gm}{\gamma}

\newcommand{\Tr}[1]{{\rm Tr}\TB{#1}}

\renewcommand{\ap}{a^\prime}
\newcommand{\bp}{b^\prime}

\newcommand{\mS}{\mathcal{S}}

\newcommand{\threeint}[1]{\int \frac{d^3 {#1}}{(2\pi)^3}}

\newcommand{\Pp}{\mathds{P}_+}

\newcommand{\Pm}{\mathds{P}_-}
\newcommand{\Ppm}{\mathds{P}_{\pm}}

\newcommand{\Sg}{\Sigma}

\newcommand{\FSumInt}[1]{\SumInt_{\SB{#1}}}
\newcommand{\nB}{n_B}
\newcommand{\nFRp}{n_F^{R,+}}

\newcommand{\nFLp}{n_F^{L,+}}

\newcommand{\spr}{s^{\prime}}

\newcommand{\nFRpm}{n_F^{R,\pm}}
\newcommand{\nFLpm}{n_F^{L,\pm}}

\allowdisplaybreaks

\begin{document}
	\title{Plasminos in chiral QCD plasma}
			
	\author{Sourav Duari\orcidlink{0009-0006-0795-5186}$^{a,c}$}
	\email{s.duari@vecc.gov.in}
	\email{sduari.vecc@gmail.com}
	
	\author{Nilanjan Chaudhuri\orcidlink{0000-0002-7776-3503}$^{a}$}
	\email{n.chaudhri@vecc.gov.in}
	\email{nilanjan.vecc@gmail.com}

	\author{Sourav Sarkar\orcidlink{0000-0002-2952-3767}$^{a,c}$}
	\email{sourav@vecc.gov.in}
	
	\author{Pradip Roy\orcidlink{0009-0002-7233-4408}$^{b,c}$}
	\email{pradipk.roy@saha.ac.in}

	\affiliation{$^a$Variable Energy Cyclotron Centre, 1/AF Bidhannagar, Kolkata - 700064, India}
	\affiliation{$^b$Saha Institute of Nuclear Physics, 1/AF Bidhannagar, Kolkata - 700064, India}
	\affiliation{$^c$Homi Bhabha National Institute, Training School Complex, Anushaktinagar, Mumbai - 400085, India}
%
	
	\begin{abstract}
The quark self-energy in a hot and dense QCD medium with local chiral imbalance shows additional structures. Evaluated using the hard thermal loop approximation this leads to distinct dispersion relations for left and right handed  quark quasi-particle and plasmino modes.

	\end{abstract}
	
	\maketitle

\section{Introduction}

The study of novel transport phenomena in the presence of chiral fermions has gained significant attention in recent years with implications spanning from high-energy nuclear physics to condensed matter systems~\cite{Vilenkin:1980fu,Kharzeev:2013ffa,Miransky:2015ava,Kharzeev:2022ydx}. A particularly well-studied example is the Chiral Magnetic Effect (CME)~\cite{Kharzeev:2007jp,Fukushima:2008xe,Kharzeev:2012ph,Kharzeev:2015znc,Koch:2016pzl} which arises from an interplay between chiral imbalance and external magnetic fields. The CME offers a promising avenue to probe topological fluctuations in Quantum Chromodynamics (QCD)~\cite{Kharzeev:2020jxw} and similar mechanisms have been explored in Dirac and Weyl semimetals~\cite{Son:2012bg,Gorbar:2013dha,Li:2014bha,Cortijo:2016wnf,Kaushik:2018tjj,Sukhachov:2021fkh} where chiral quasiparticles are realized in condensed matter systems.

In QCD, the non-Abelian gauge structure of the $SU(3)$ group leads to a rich topological vacuum structure, comprising an infinite number of physically equivalent configurations whose vacua are topologically  distinct. These vacuum configurations of QCD at zero and low temperatures can be characterised by topologically non-trivial gauge configurations with a non-zero winding number~\cite{Shifman:1988zk,Lenz:2001me}. These gluon field configurations, known as instantons,  mediate transitions between distinct vacua by tunneling through a potential barrier whose height is of the order of the QCD scale, $\Lambda_{\rm QCD}$. This non-perturbative process is referred to as instanton tunneling~\cite{Belavin:1975fg,tHooft:1976rip,tHooft:1976snw}. However, at high temperature such as those realized in the quark-gluon plasma (QGP) created during heavy-ion collisions, a substantial production of another class of gluon configurations, known as sphalerons, is expected~\cite{Manton:1983nd,Klinkhamer:1984di}. It is conjectured that the high abundance of sphalerons enable real-time thermally activated transitions over the energy barrier separating topologically distinct vacua~\cite{Kuzmin:1985mm,Arnold:1987mh,Khlebnikov:1988sr,Arnold:1987zg}. These topologically nontrivial gauge field configurations can flip the helicities of quarks through their interactions, thereby generating an imbalance between left- and right-handed quarks. This process, a consequence of the axial anomaly of QCD, results in local violations of parity ($P$) and charge-parity ($CP$) symmetries~\cite{Adler:1969gk,Bell:1969ts}. Although no global \(CP\) violation has been observed in strong interactions, such local violations may lead to domains of finite chiral imbalance in the QGP~\cite{McLerran:1990de,Moore:2010jd}. This imbalance is commonly characterized by a chiral chemical potential \( \mu_5 \), which quantifies the difference in number densities between right- and left-handed quarks.

In the presence of a chiral imbalance, it becomes essential to re-examine the collective behavior of the medium. The spectrum of collective excitations encodes key information about the thermodynamic and transport properties of the system, both in and out of equilibrium. Particle propagation in such a medium leads to interaction-induced modifications including effective mass generation and the emergence of collective modes with dispersion relations that differ markedly from those in vacuum. To systematically account for these medium-induced modifications, one must employ resummed perturbative techniques specifically designed for thermal gauge theories such as QCD and QED, where naive perturbation theory fails due to infrared divergences arising from massless gauge bosons. This issue is addressed through the hard thermal loop (HTL) resummation program~\cite{Braaten:1989mz,Andersen:1999fw}, which reorganizes the perturbative expansion around effective quasiparticles endowed with thermal masses. In the gluonic sector, it has been shown that chiral imbalance can induce instabilities even in a spatially isotropic plasma, particularly at momenta much smaller than the Debye mass. The presence of a chiral chemical potential also leads to a splitting of the transverse gluon modes into left and right-handed circularly polarized branches. Furthermore, the characteristic screening length of the medium, the Debye radius, has been found to decrease with increasing chiral chemical potential, suggesting a stronger screening effect and consequently an enhanced suppression of quarkonium states~\cite{Akamatsu:2013pjd,Carignano:2018thu,Carrington:2021bnk,Duari:2025kxt}.

In chirally symmetric media, the fermionic spectrum features two distinct branches: a particle-like mode and a collective hole-like excitation known as the plasmino~\cite{Klimov:1981ka,Weldon:1982bn,Weldon:1989bg,Weldon:1989ys,Bellac:2011kqa,Strickland:2019tnd,Mustafa:2022got}. These correspond to time-like poles of the thermal fermion propagator and reflect the nontrivial structure of the medium. The plasmino, absent at zero temperature, arises purely from thermal effects and exhibits a non-monotonic dispersion with a minimum at finite momentum, a feature tied to the collective nature of the excitation.

In this work, we present a systematic investigation of quark collective excitations in a hot and chirally imbalanced QCD medium, an aspect that remains largely unexplored in the existing literature. We begin by analyzing the general Dirac structure of the fermion self-energy in the presence of a chiral chemical potential. We then compute the one-loop fermion self-energy using the HTL approximation, carefully identifying the nontrivial structure functions that distinguish left- and right-handed sectors. Using the Schwinger–Dyson approach, we construct the effective fermion propagator and study the dispersion relations of the resulting excitations. Special attention is given to the interplay between helicity and chirality, which governs the character of the collective modes. The resulting propagator can also serve as an input for evaluating observables such as photon damping rate, real photon emission and lepton pair production from chirally asymmetric media.

\section{General structure of fermion self-energy in presence of chiral imbalance}
In this section we discuss the general structure of the self-energy of a massless fermion in a chially imbalanced medium. Note that the fermion self-energy is a $ 4\times 4 $ matrix in the Dirac space and it is a Lorentz scalar. At finite temperature and in the presence of chiral imbalance it is a function of four-momentum $ P $ of the corresponding fermion and medium four-velocity $ U $. It is well known that any $ 4\times 4 $ matrix can be expressed in terms of sixteen basis matrices:$ \SB{\unit, \gm_5,\gm^\mu, \gm^\mu\gm_5,\sigma^{\munu}} $ where $ \gm_5  = i\gm^0\gm^1\gm^2\gm^3$ and $ \sigma^\munu = i/2 \TB{\gm^\mu,\gm^\nu} $. Terms involving $ \sigma^\munu $ will not arise due to its anti-symmetric nature in any loop order in the self-energy.  We denote all four vectors by capital letters i.e. $P^\mu = \FB{p_0, \vec{p}}$ and our metric is mostly negative. Thus, for a massless fermion we arrive at the following general covariant structure of the fermion self-energy
\begin{equation}\label{Eq_GenS}
	\Sigma(P) = - a \slashed{P} -b\slashed{U} - \ap \gm_5 \slashed{P}- \bp \gm_5 \slashed{U}~.
\end{equation} 
The structure functions \( a \) and \( b \) are present even in a parity-symmetric thermal medium, while the additional functions \( a' \) and \( b' \) arise solely due to the presence of a chiral imbalance. Note that, terms proportional to $\gm_5 $ do not contribute at one loop level. By projecting Eq.~\eqref{Eq_GenS} onto appropriate basis structures and subsequently taking the trace, all the form factors can be systematically extracted from the self-energy \( \Sigma(P) \). The resulting expressions are given below:

\begin{align}
	a &= \dfrac{1}{4 p^2}~\FB{\Tr{\fsl{P}\Sg } -\FB{P\cdot U} \Tr{\slashed{U}\Sg \textcolor{white}{\dfrac{}{}}} },\label{Eq_a} \\
	b &= \dfrac{1}{4 p^2}~\FB{-\FB{P\cdot U}\Tr{\fsl{P}\Sg } + P^2 \Tr{\slashed{U}\Sg} \textcolor{white}{\dfrac{}{}}}, \label{Eq_b}\\
	\ap &= -\dfrac{1}{4p^2}~\FB{~\Tr{\gm_5\fsl{P}\Sg } - \FB{P\cdot U} \Tr{\gm_5\slashed{u}\Sg} \textcolor{white}{\dfrac{}{}}}~, \label{Eq_ap}\\
	\bp &= -\dfrac{1}{4p^2}~\FB{ -\FB{P\cdot U}\Tr{\gm_5\fsl{P}\Sg } + P^2 \Tr{\gm_5\slashed{U}\Sg}  \textcolor{white}{\dfrac{}{}}}~.\label{Eq_bp}
\end{align}

\section{One loop self energy of a quark}
The quark propagator in presence of chiral imbalance consists of two parts describing the modes with right (+) and left(-) handed chirality and is given by~\cite{Kharzeev:2009pj,Ghosh:2022xbf}
\begin{equation}
	S(K) = \Pp \dfrac{\slashed{K}_+}{K_+^2} + \Pm \dfrac{\slashed{K}_-}{K_-^2}= \sum_{s\in \pm} \mathds{P}_s \dfrac{\slashed{K}_s}{K_s^2}~,\label{Eq_propF_mu5}
\end{equation}
\begin{figure}[h]
	\centering
	\includegraphics[scale=0.13]{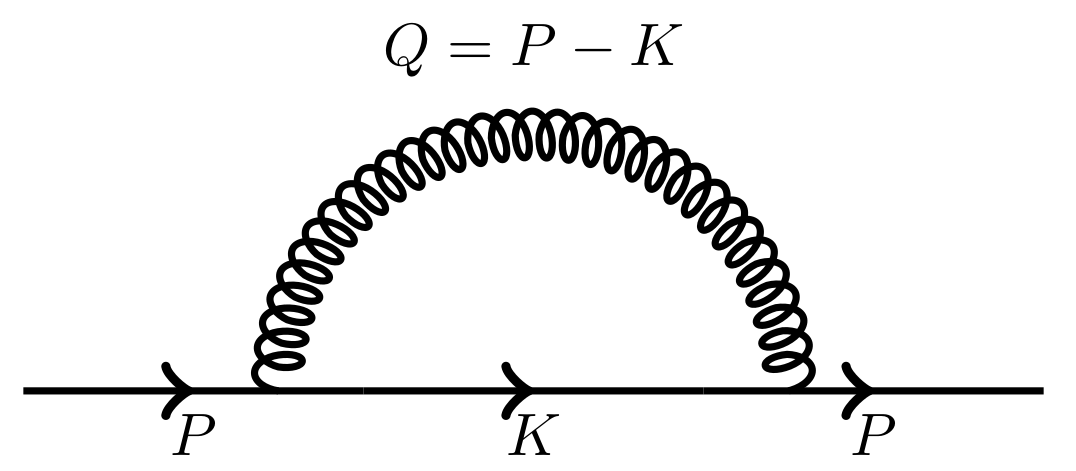}
	\caption{one loop quark self energy.}
	\label{Fig_self_energy}
\end{figure}
where $\Ppm = \frac{1}{2} (1 \pm \gm_5)$ are chirality projection operators and $K^\mu_\pm = \FB{k_0+\mu \mp \mu_5, \vec{k}} = \FB{k_0+\mu_\pm, \vec{k}}= \FB{k^\pm_0, \vec{k}}$ with $\mu_\pm = \mu \mp \mu_5$. Accordingly $K^2_\pm = \FB{k^\pm_0}^2 -\vec{k}^2$. As we are dealing with massless fermions the opposite chiralities do not mix.

Now, using the imaginary time formulation of thermal field theory, the quark self-energy in the Feynman gauge can be obtained from the Feynman diagram in Fig.~\ref{Fig_self_energy} as
\begin{align}
	\Sigma(P) &= g^2 C_F \SumInt_{\SB{K}} \gm_\mu \FB{\Pp \dfrac{\slashed{K}_+}{K^2_+} + \dfrac{\slashed{K}_-}{K^2_-}}\gm^\mu \frac{1}{Q^2}\nn \\ 
	&= - 2g^2 C_F \SumInt_{\SB{K}} \FB{\Pp \dfrac{\slashed{K}_+}{K^2_+} + \dfrac{\slashed{K}_-}{K^2_-}} \frac{1}{Q^2}\label{Eq_self_energy}
\end{align}
where $Q = P-K$ and $\FSumInt{K} = T \sum_{k_0}\threeint{k}$ is fermionic sum-integral with $k_0 = i \frac{(2n+1)\pi}{\beta}$. We now  compute the form factors appearing in the general structure of the self-energy as expressed in Eq.~\eqref{Eq_GenS}. Starting from Eq.~\eqref{Eq_self_energy} we arrive at the  following results  
\begin{align}
	\Tr{\slashed{P}\Sg} &=- 4 g^2 C_F \FSumInt{K} \TB{\dfrac{k_0^+ p_0 -\vec{p}\cdot\vec{k}}{K^2_+}+\dfrac{k_0^- p_0 -\vec{p}\cdot\vec{k}}{K^2_-} } \frac{1}{Q^2} \label{Eq_trslPSg}\\
	\Tr{\slashed{U}\Sg} &=- 4 g^2 C_F \FSumInt{K} \TB{\dfrac{k_0^+ }{K^2_+}+\dfrac{k_0^-}{K^2_-} } \frac{1}{Q^2}\\
	\Tr{\gm_5\slashed{P}\Sg} &=- 4 g^2 C_F \FSumInt{K} \TB{-\dfrac{k_0^+ p_0 -\vec{p}\cdot\vec{k}}{K^2_+}+\dfrac{k_0^- p_0 -\vec{p}\cdot\vec{k}}{K^2_-} } \frac{1}{Q^2} \\
	\Tr{\gm_5\slashed{U}\Sg} &=- 4 g^2 C_F \FSumInt{K} \TB{-\dfrac{k_0^+ }{K^2_+}+\dfrac{k_0^-}{K^2_-} } \frac{1}{Q^2}~.\label{Eq_trgm5Sg}
\end{align}
The required Matsubara sums are provided in Appendix~\ref{Sec_App}. Using Eqs.\eqref{Eq_a}–\eqref{Eq_bp}, one can subsequently determine all the relevant form factors. In what follows, we focus on computing the nontrivial structure functions appearing in Eq.\eqref{Eq_GenS} within the framework of the hard thermal loop (HTL) approximation. This approach captures the leading-order thermal contributions in the high temperature regime. In this approximation, the loop momentum $K$ is considered hard, i.e. of order $T$, while the external momentum $P$ is treated as soft i.e. of order $gT \ll T$. A detailed overview of the HTL approximation and its physical implications can be found in Refs.~\cite{Klimov:1982bv,Weldon:1982bn,Braaten:1989mz,Bellac:2011kqa,Laine:2016hma,Strickland:2019tnd,Mustafa:2022got,Haque:2024gva}. Within the framework of the HTL approximation, a number of systematic simplifications arise owing to the well-defined separation between hard and soft momentum scales. These are outlined as follows
\begin{align}
	q = \MB{\vec{p}-\vec{k}}&= k-p \cos \theta = k-\vec{p}\cdot\hat{k}
	\nB (q) \\
	\nB (k-\vec{p}\cdot\hat{k}) &\simeq \nB (k) - \vec{p}\cdot\hat{k} \dfrac{d \nB (k)}{d k }\\
	p_0\pm k\pm q &\simeq p_0 \pm k \pm k \mp \vec{p}\cdot\hat{k}\simeq \pm 2 k\\
	p_0\pm k\mp q &\simeq p_0 \pm k \mp k \pm \vec{p}\cdot\hat{k}\simeq p_0 \pm \vec{p}\cdot\hat{k}~.
\end{align}
Employing these approximations in Eqs.~\eqref{Eq_trslPSg} to \eqref{Eq_trgm5Sg} one arrives at the following expressions of the structure factors in HTL approximation
\begin{align}
	a &= - \dfrac{M^2}{p^2} Q_1(p_0, p)  \\
	b &= \dfrac{M^2}{p} \TB{\frac{p_0}{p} Q_1(p_0,p) - Q_0 (p_0,p)  }\\
	\ap &=  \dfrac{\delta M^2}{p^2} Q_1(p_0, p)\\
	\bp &= \dfrac{\delta M^2}{p} \TB{\frac{p_0}{p} Q_1(p_0,p) -  Q_0 (p_0,p)  }
\end{align}
where 
\begin{align}
	M^2 &= M^2 (T,\mu,\mu_5) = \frac{g^2 C_F}{8} \FB{T^2+ \frac{\mu^2}{\pi^2}+ \frac{\mu_5^2}{\pi^2}} \\
	\delta M^2 &= \frac{g^2 C_F}{4\pi^2} \mu \mu_5~~~~~
\end{align}
and $Q_0$ and $Q_1$ are Legendre functions of second kind and defined as
\begin{equation}
	Q_0 (p_0,p) = \frac{1}{2} \ln \TB{\frac{p_0+p}{p_0-p}}~~~~~~~~~~~~~~	
	Q_1 (p_0,p) = \frac{p_0}{p} Q_0(p_0,p) -1~.
\end{equation}
Thermal integrals used to arrive at these results are given in Appendix~\ref{sec_thermal_int}. Here $M$ denotes the thermal mass of the quark while $\delta M $ represents an additional mass scale that arises due to the presence of chiral asymmetry. Since $\delta M$ is proportional to $\mu_5$ it is straightforward to verify that in the limit $\mu,\mu_5 \to 0$ the structure factors $\ap$ and $\bp$ vanish and the results reduce to the standard HTL expressions for a  thermal medium, as established in Refs.~\cite{Klimov:1982bv,Weldon:1982bn,Braaten:1989mz,Bellac:2011kqa,Laine:2016hma,Strickland:2019tnd,Mustafa:2022got,Haque:2024gva}.

\section{Collective excitation of quarks in chirally imbalanced media}

Employing Dyson-Schwinger equation the inverse of the effective fermion propagator can be expressed as
\begin{equation}\label{Eq_eff_prop}
	\mS^{-1} (P) = \slashed{P}- \Sg(P)=\Pp \slashed{L} \Pm + \Pm \slashed{R} \Pp
\end{equation}
with 
\begin{align}
	\slashed{L} &= (1+a +\ap ) \slashed{P} +(b+\bp) \slashed{U}~,\\
	\slashed{R} &= (1+a -\ap ) \slashed{P} +(b-\bp) \slashed{U}~.
\end{align}
Then the propagator itself becomes~\cite{Weldon:1982bn}
\begin{equation}
	\mS(P) = \slashed{P}- \Sg(P)=\Pp \dfrac{\slashed{L}}{L^2} \Pm + \Pm \frac{\slashed{R}}{R^2} \Pp \label{Eq_full_prop}
\end{equation}
with 
\begin{align}
	L^2 &= \SB{\FB{1+a +\ap} p_0 +\FB{b+\bp} }^2-\SB{\FB{1+a +\ap} p }^2~, \\
	R^2 &= \SB{\FB{1+a -\ap} p_0 +\FB{b-\bp} }^2-\SB{\FB{1+a -\ap} p }^2~.
\end{align}
The dispersion properties of quarks in a chirally imbalanced medium are governed by the zeros of $L^2$ and $R^2$ which we refer to as the L-mode and R-mode, respectively. To establish a baseline, we begin by considering the case when $\mu,\mu_5 = 0$. In this limit, the structure factors $\ap$ and $\bp$ vanish identically resulting in a degeneracy between $L$ and $R$-modes. The dispersion relation for quark excitations in a thermal medium is then obtained from the condition
\begin{equation}
	\SB{\FB{1+a } p_0 +b }^2-\SB{\FB{1+a } p }^2 = 0~.
\end{equation}  
The qualitative features of the resulting quark dispersion curves in the thermal medium can be interpreted following~\cite{Bellac:2011kqa}. 
At zero temperature, massless fermions exhibit chiral symmetry and possess a strict correlation between the sign of energy and chirality-to-helicity ratio $\chi$: positive (negative) frequency modes have $\chi=+1 (\chi=-1)$. Thermal effects, however, modify this structure as both positive  and negative chirality-to-helicity modes acquire time-like poles signifying well-defined quasiparticles. At vanishing momentum, both modes acquire a common thermal mass $M (T,0,0)$. 
At large momentum, the positive branch behaves like a free particle with an asymptotic thermal mass correction while the negative branch, known as the plasmino, exhibits a non-monotonic dispersion and an exponentially suppressed residue.
The existence of plasmino mode is purely a phenomena induced by the thermal medium.

\begin{figure}[h]
	\includegraphics[scale=0.25]{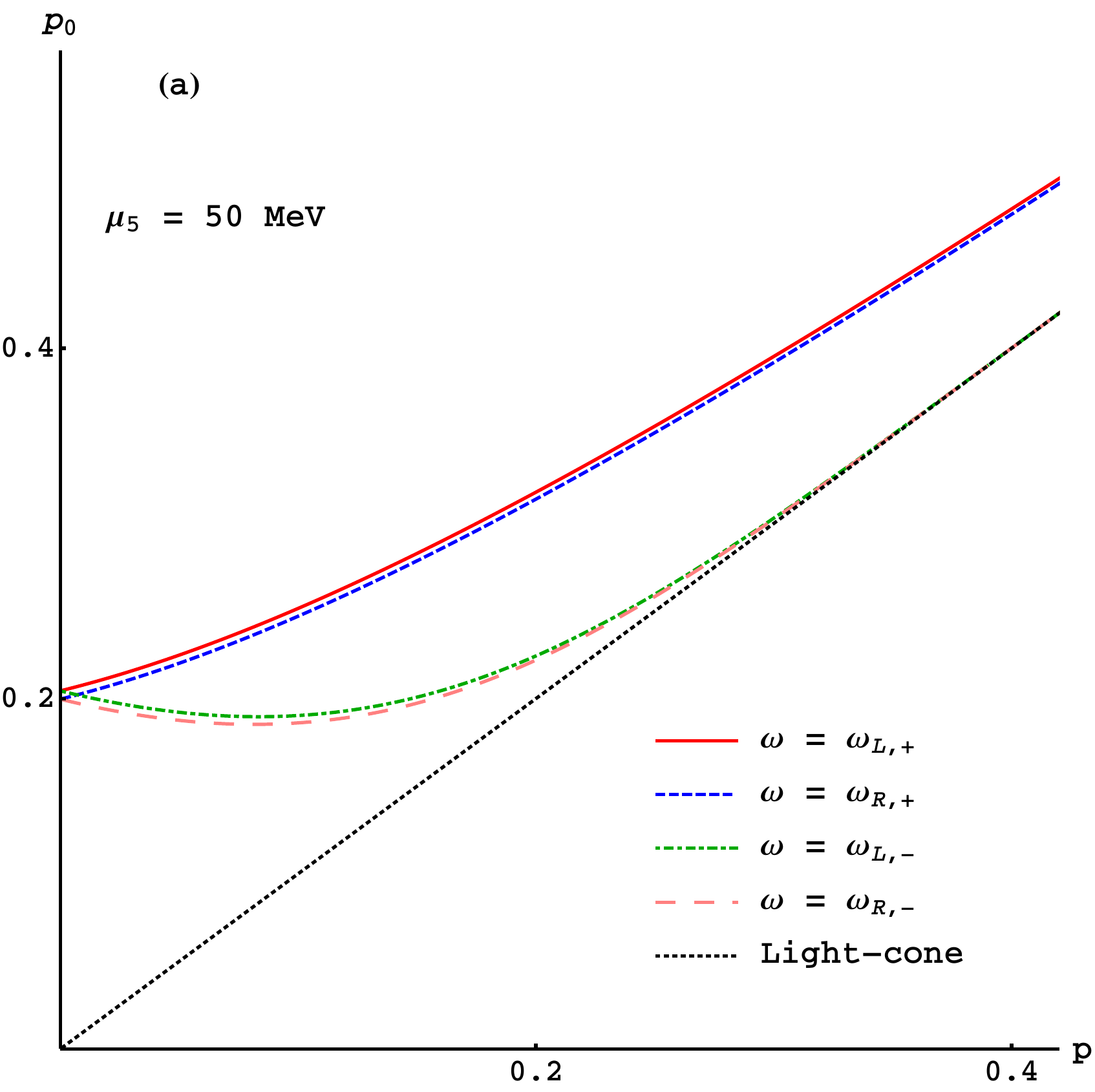} ~~~~~~~
	\includegraphics[scale=0.25]{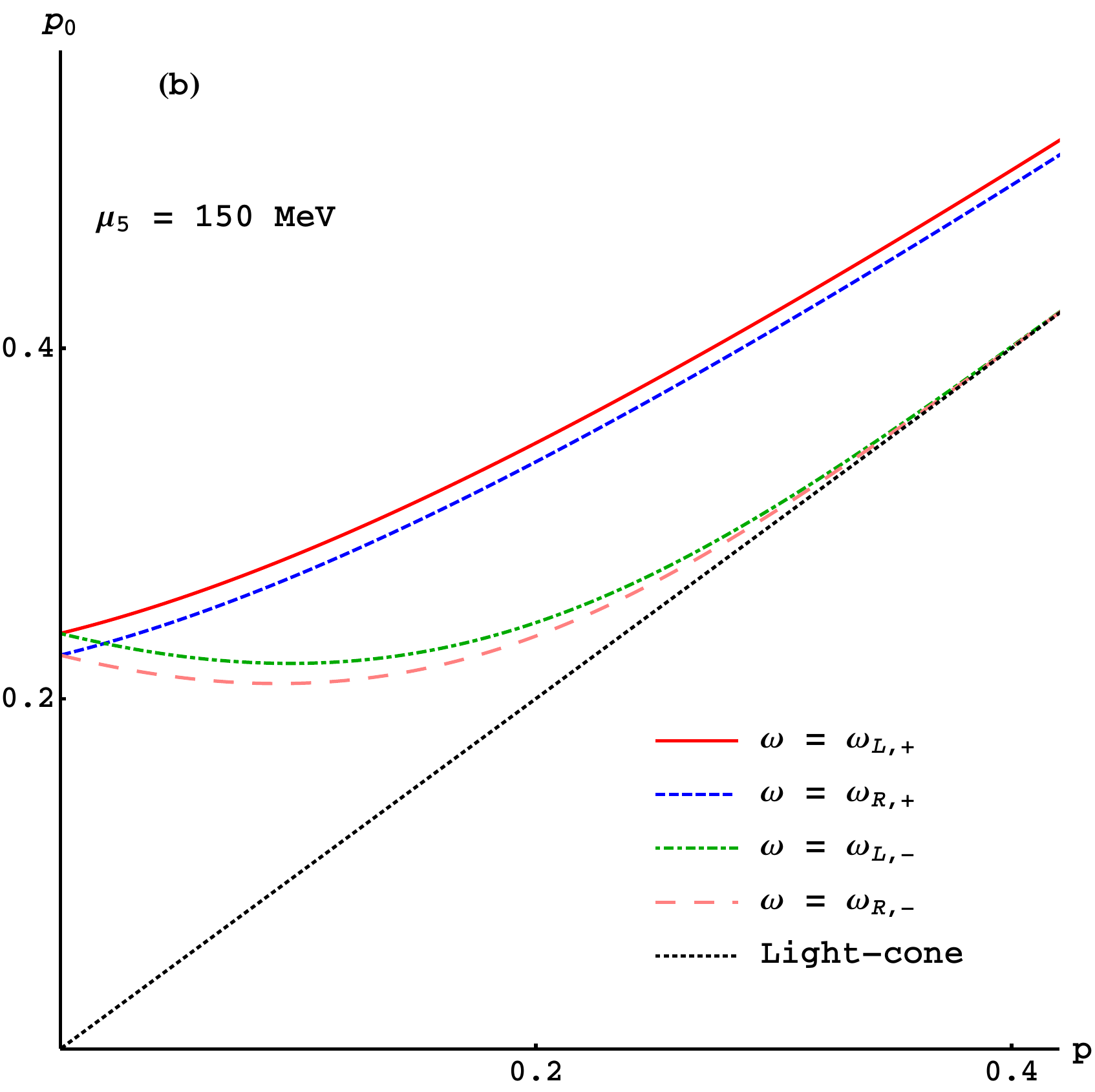}
	\caption{Dispersion of $L$ and $R$-mode for different values of $\mu_5$ at $T=200$ MeV and $\mu = 150$ MeV.}
	\label{Fig_Dispersion}
\end{figure}

Now the presence of $\mu_5$ does not break chiral invariance as $\mS^{-1} (P)$  remains invariant under chiral transformations and anti-commutes with $\gamma_5$ by construction. Moreover  $\mS^{-1} (P)$ also anti-commutes with the helicty operator $\frac{\gm_0 \vec{\gm}\cdot \vec{p}}{p}$. Thus these spinors which satisfy the modified Dirac 	equation in chirally imbalanced medium must be simultaneous eigenstates of both helicity and chirality. In a chiral plasma  $L$ and $R$-modes become non-degenerate due to local parity violation and each has two physical (positive energy) solutions which are plotted in Fig.~\ref{Fig_Dispersion}.  Both  $L$ and $R$-modes possess time-like poles with positive or negative chirality-to-helicity ratios. 
The interplay between helicity and chirality becomes transparent by recasting the full propagator, as given in Eq.~\eqref{Eq_full_prop}, in the following form~\cite{Blaizot:2001nr}
\begin{align}
	\mS (P) &= \Pp \FB{\dfrac{\gm_0\Lambda_+}{ A^L_0 -A^L_s}+\dfrac{\gm_0\Lambda_-}{ A^L_0 +A^L_s}}\Pm \nn \\ 
	&~~~~~~+ \Pm \FB{\dfrac{\gm_0\Lambda_+}{ A^R_0 -A^R_s}+\dfrac{\gm_0\Lambda_-}{ A^R_0 +A^R_s}}\Pp
\end{align}
where 
\begin{align}
	\Lambda_\pm &= \half \FB{1 \mp \gm_0 \frac{\vec{\gm} \cdot \vec p}{p}}~,\\
	A^{L,R}_0 &= \FB{1+a\pm\ap} p_0  + \FB{b\pm\bp}~, \\
	A^{L,R}_s &= \FB{1+a\pm\ap} p~. 
\end{align}
The operators $\Lambda_\pm$ defined above  projects out spinors whose chirality is equal or opposite to the helicity. Accordingly, for $L$-modes, the pole of $ A^L_0 -A^L_s = 0$ is interpreted as particle-like solution and the pole of $ A^L_0 + A^L_s = 0$ gives rise to the hole-like excitation, commonly referred to as the plasmino. We have denoted them as $\omega_{L,+}$ and $\omega_{L,-}$ respectively. An analogous interpretation holds for the dispersion of $R$-modes. 

Numerical solutions for two representative values of $\mu_5$ are shown in Figs.~\ref{Fig_Dispersion} (a) and (b) illustrating the dispersion relations for both $L$ and $R$-modes. As can be seen, in the chiral plasma $L$ and $R$-modes disperse differently. At vanishing momentum particle and plasmino solutions of $L$ and $R$-modes acquire different thermal mass $M_+$ and $M_- $  respectively with $M_\pm  = M \pm \delta M $. These are the fermionic analogues of the plasma frequency. The splitting between the effective thermal masses of  $L$ and $R$-modes increase with $\mu_5$ as $\delta M$ is proportional to $\mu_5$. This trend is clearly visible when one compares Figs.~\ref{Fig_Dispersion} (a) and (b) in the limit $p\to 0$. 
At intermediate momenta, the plasmino branches of both $L$ and $R$-modes exhibit a characteristic non-monotonic dispersion. These modes can be interpreted as collective excitations arising from the mixing between positive-frequency fermion states and negative-frequency anti-fermion states induced by thermal effects. 
In particular, at small momentum the dispersion of both the plasmino branches $\omega_{L,-}$ and $\omega_{R,-}$ behave similar to a negative-energy mode as the energy decreases with increasing momentum. These features can be captured analytically through approximate solutions valid in the limit $p \ll M_\pm$. These solutions are given by
\begin{align}
	\omega_{L,+} &\simeq M_+ + \frac{1}{3}p+ \frac{1}{3 M_+ } p^2 +\cdots \label{Eq_lowp_part}\\
	\omega_{L,-} &\simeq M_+  - \frac{1}{3} p+ \frac{1}{3 M_+ } p^2 +\cdots \label{Eq_lowp_hole}
\end{align}
A similar expression holds for the $R$-mode obtained by replacing  $M_+ $ with $M_-$.
At asymptotic momenta $ p \gg M_\pm$, both $\omega_{L,+}$ and $\omega_{R,+}$ clearly develop an asymptotic mass $ M_+$ and $M_-$ respectively, while plasmino solutions $\omega_{L,-}$ and $\omega_{R,-}$ with negative helicity to chirality ratio approach the light cone exponentially fast. This is evident from the following behaviour of the particle and hole-like solutions at large momenta $(M_\pm \ll p \ll T)$
\begin{align}
	\omega_{L,+} &\simeq p + \frac{M^2_+}{p}  +   \dfrac{M_+^4}{2 p^3} \ln \frac{M_+^2}{2 p^2} \cdots\\
	\omega_{L,-} &\simeq p + 2p \exp\TB{-\frac{M_+^2 + 2p^2}{M^2_+}}\label{Eq_highp_hole}~.
\end{align}
Similar expressions can be written for the $R$-mode with the substitution $M_+ \to M_-$.

A distinctive feature of the plasmino branch is the presence of a minimum at finite momentum. This non-monotonic behavior can be understood as follows. At $p =0$, the slope of the dispersion curve for the hole-like excitation is negative indicating that the group velocity is oppositely directed to the momentum. In contrast, for the particle-like excitation the group velocity has the same sign as the momentum. This behavior is evident from Eqs.\eqref{Eq_lowp_part} and \eqref{Eq_lowp_hole}. At large momentum, however, the plasmino branch asymptotically approaches the light cone  with the group velocity tending toward unity as shown in Eq.\eqref{Eq_highp_hole}. The combination of these two limiting behaviors leads to a local minimum in the dispersion relation at finite momentum. The existence of such a minimum implies a vanishing group velocity at that point which can result in van Hove singularities in the density of states. These singularities in turn produce observable effects such as gaps or enhancements in physical quantities such as the di-lepton production rate~\cite{Braaten:1990wp}.

\section{Summary}
To summarise, we write down the most general structure of the fermion self-energy in a thermal medium with chiral imbalance characterized by a chiral chemical potential $\mu_5$. 
The one loop quark self-energy evaluated using the HTL approximation shows two additional structure functions compared to the case when $\mu_5 = 0$.  We observe distinct collective modes for left and right handed quark quasi-particles and holes (plasmino). Analytical expressions for different modes are obtained for small and large momenta. Finally, it is important to note that the effective  propagator obtained here is essential for evaluation of photon damping rate, real photon emission and lepton pair production from matter with chiral imbalance.

\appendix

\section{Frequency sum}\label{Sec_App}
To simplify Eqs.~\eqref{Eq_trslPSg} to \eqref{Eq_trgm5Sg} following frequency sums are necessary~\cite{Bellac:2011kqa,Mustafa:2022got,Haque:2024gva}
\begin{align}
	T\sum_{k_0}\dfrac{1}{K^2_+ Q^2} &= - \sum_{s, \spr \in \pm}\dfrac{s \spr}{4 k q} \dfrac{1- \nFRp (s k) +\nB (\spr q)} {p_0 - s k - \spr q}  \\
	T\sum_{k_0}\dfrac{k_0^+}{K^2_+ Q^2} &= - \sum_{s, \spr \in \pm}\dfrac{ \spr}{4 q} \dfrac{1- \nFRp (s k) +\nB (\spr q)} {p_0 - s k - \spr q}\\
	T\sum_{k_0}\dfrac{1}{K^2_- Q^2} &=  - \sum_{s, \spr \in \pm}\dfrac{s \spr}{4 k q} \dfrac{1- \nFLp (s k) +\nB (\spr q)} {p_0 - s k - \spr q}\\
	T\sum_{k_0}\dfrac{k_0^-}{K^2_- Q^2} &=- \sum_{s, \spr \in \pm}\dfrac{ \spr}{4 q} \dfrac{1- \nFLp (s k) +\nB (\spr q)} {p_0 - s k - \spr q}
\end{align}
with $k_0^\pm = \frac{i (2n +1) \pi}{\beta} +\mu_\pm $.  The distribution functions are defined as follows
\begin{align}
	\nFRpm (k) &= \dfrac{1}{e^{\beta (k\mp \mu_+)} +1}~,\\
	\nFLpm (k) &= \dfrac{1}{e^{\beta (k\mp \mu_-)} +1}~, \\
	\nB(q) &= \dfrac{1}{e^{\beta q} -1}~.
\end{align}

\section{Thermal integrals}\label{sec_thermal_int}
Relevant thermal integrals required in the evaluation of structure factors are~\cite{Bellac:2011kqa,Loganayagam:2012pz}
\begin{align}
	\int k ~dk~ \nB (k) &= \frac{\pi^2 T^2}{6}~, \\
	\int k ~dk ~\FB{ n_F^+(k)+ n_F^-(k) } &=  \frac{\pi^2 T^2}{6}+ \frac{\mu^2}{2} ~,\\
	\int  dk \FB{ n_F^+(k)- n_F^-(k) }&= \mu ~.
\end{align}

\bibliography{reference} 

	\end{document}